\documentclass[%
 reprint,
 amsmath,amssymb,
 aps,prl
]{revtex4-2}

\usepackage{graphicx}% Include figure files
\usepackage{dcolumn}% Align table columns on decimal point
\usepackage{bm}% bold math
\usepackage[mathlines]{lineno}% Enable numbering of text and display math
\usepackage{booktabs}

\begin{document}

%\preprint{APS/123-QED}

\title{Search for reactor-produced millicharged particles with Skipper-CCDs\\ at the CONNIE and Atucha-II experiments}

\author{Alexis A.     Aguilar-Arevalo}
\affiliation{Instituto de Ciencias Nucleares, Universidad Nacional Autónoma de México,  Circuito Exterior s/n, C.U., CDMX, Mexico}

\author{Nicolas       Avalos}
\affiliation{Centro Atómico Bariloche and Instituto Balseiro, Comisión Nacional de Energía Atómica (CNEA), Consejo Nacional de Investigaciones Científicas y Técnicas (CONICET), Universidad Nacional de Cuyo (UNCUYO), Av Bustillo 9500, San Carlos de Bariloche, Argentina}

\author{Pablo Bellino}
\affiliation{Comisión Nacional de Energía Atómica, Centro Atómico Constituyentes} 

\author{Xavier        Bertou}
\affiliation{Centro Atómico Bariloche and Instituto Balseiro, Comisión Nacional de Energía Atómica (CNEA), Consejo Nacional de Investigaciones Científicas y Técnicas (CONICET), Universidad Nacional de Cuyo (UNCUYO), Av Bustillo 9500, San Carlos de Bariloche, Argentina}

\author{Carla       Bonifazi}
\affiliation{International Center for Advanced Studies \& Instituto de Ciencias Físicas, ECyT-UNSAM and CONICET, 25 de Mayo y Francia, Buenos Aires, Argentina}
\affiliation{Instituto de Física, Universidade Federal do Rio de Janeiro, Av. Athos da Silveira Ramos, 149, Cidade Universitária, Rio de Janeiro, RJ, Brazil}

\author{Ana Botti}
\affiliation{Fermi National Accelerator Laboratory, Batavia, IL, United States} 

\author{Mariano Cababié}
\affiliation{Institut für Hochenergiephysik der Österreichischen Akademie der Wissenschaften, 1050 Wien - Austria}
\affiliation{Atominstitut, Technische Universität Wien, 1020 Wien - Austria}

\author{Gustavo       Cancelo}
\affiliation{Fermi National Accelerator Laboratory, Batavia, IL, United States}

\author{Brenda A.   Cervantes-Vergara}
\affiliation{Instituto de Ciencias Nucleares, Universidad Nacional Autónoma de México,  Circuito Exterior s/n, C.U., CDMX, Mexico}
\affiliation{Fermi National Accelerator Laboratory, Batavia, IL, United States}

\author{Claudio  Chavez}
\affiliation{Fermi National Accelerator Laboratory, Batavia, IL, United States} 
\affiliation{Instituto de Inv. en Ing. Eléctrica “Alfredo Desages” (IIIE), Dpto. de Ing. Eléctrica y de Computadoras, CONICET and Universidad Nacional del Sur (UNS), 800 San Andrés Street, Bahía Blanca, Argentina} 
\affiliation{Facultad de Ingeniería, Universidad Nacional de Asunción, Campus de la UNA, San Lorenzo, Paraguay}

\author{Fernando      Chierchie}
\affiliation{Instituto de Inv. en Ing. Eléctrica “Alfredo Desages” (IIIE), Dpto. de Ing. Eléctrica y de Computadoras, CONICET and Universidad Nacional del Sur (UNS), 800 San Andrés Street, Bahía Blanca, Argentina}

\author{David Delgado}
\affiliation{Central Nuclear ATUCHA I-II, Nucleoeléctrica Argentina Sociedad Anónima, Buenos Aires, Argentina} 

\author{Eliana Depaoli}
\affiliation{Comisión Nacional de Energía Atómica, Centro Atómico Constituyentes}%
\affiliation{Universidad de Buenos Aires, Facultad de Ciencias Exactas y Naturales, Departamento de Física, Buenos Aires, Argentina}

\author{Juan Carlos   D'Olivo}
\affiliation{Instituto de Ciencias Nucleares, Universidad Nacional Autónoma de México,  Circuito Exterior s/n, C.U., CDMX, Mexico}

\author{João          dos Anjos}
\affiliation{Centro Brasileiro de Pesquisas Físicas, Rua Dr. Xavier Sigaud, 150, Urca,  Rio de Janeiro, RJ, Brazil}

\author{Juan          Estrada}
\affiliation{Fermi National Accelerator Laboratory, Batavia, IL, United States}

\author{Guillermo   Fernandez Moroni}
\affiliation{Fermi National Accelerator Laboratory, Batavia, IL, United States}
\affiliation{Instituto de Inv. en Ing. Eléctrica “Alfredo Desages” (IIIE), Dpto. de Ing. Eléctrica y de Computadoras, CONICET and Universidad Nacional del Sur (UNS), 800 San Andrés Street, Bahía Blanca, Argentina}
\affiliation{Department of Astronomy and Astrophysics, University of Chicago, 5640 South Ellis Avenue, Chicago, IL, 60637, USA}

\author{Aldo R.       Fernandes Neto}
\affiliation{Centro Federal de Educação Tecnológica Celso Suckow da Fonseca, Campus Angra dos Reis, Rua do Areal, 522, Pq Mambucaba, Angra dos Reis, RJ, Brazil}

\author{Richard       Ford}
\affiliation{Fermi National Accelerator Laboratory, Batavia, IL, United States}

\author{Ben           Kilminster}
\affiliation{Physik Institut, Universität Zürich, Winterthurerstrasse 190, Zurich, Switzerland}

\author{Kevin         Kuk}
\affiliation{Fermi National Accelerator Laboratory, Batavia, IL, United States}

\author{Andrew        Lathrop}
\affiliation{Fermi National Accelerator Laboratory, Batavia, IL, United States}

\author{Patrick       Lemos}
\affiliation{Instituto de Física, Universidade Federal do Rio de Janeiro, Av. Athos da Silveira Ramos, 149, Cidade Universitária, Rio de Janeiro, RJ, Brazil}

\author{Herman P.     Lima Jr.}
\affiliation{Centro Brasileiro de Pesquisas Físicas, Rua Dr. Xavier Sigaud, 150, Urca,  Rio de Janeiro, RJ, Brazil}

\author{Martin      Makler}
\affiliation{International Center for Advanced Studies \& Instituto de Ciencias Físicas, ECyT-UNSAM and CONICET, 25 de Mayo y Francia, Buenos Aires, Argentina}
\affiliation{Centro Brasileiro de Pesquisas Físicas, Rua Dr. Xavier Sigaud, 150, Urca,  Rio de Janeiro, RJ, Brazil}

\author{Agustina       Magnoni}
\affiliation{Departamento de Física, FCEN, Universidad de Buenos Aires, Buenos Aires, Argentina} 
\affiliation{Laboratorio de Óptica Cuántica, DEILAP, UNIDEF (CITEDEF-CONICET), Buenos Aires, Argentina}

\author{Katherine     Maslova}
\affiliation{Instituto de Física, Universidade Federal do Rio de Janeiro, Av. Athos da Silveira Ramos, 149, Cidade Universitária, Rio de Janeiro, RJ, Brazil}

\author{Franciole    Marinho}
\affiliation{Departamento de Física, Instituto Tecnológico de Aeronáutica, DCTA,
   12228-900, São José dos Campos, São Paulo, Brazil}
   
\author{Jorge        Molina}
\affiliation{Facultad de Ingeniería, Universidad Nacional de Asunción, Campus de la UNA, San Lorenzo, Paraguay}

\author{Irina         Nasteva}
\affiliation{Instituto de Física, Universidade Federal do Rio de Janeiro, Av. Athos da Silveira Ramos, 149, Cidade Universitária, Rio de Janeiro, RJ, Brazil}

\author{Ana Carolina  Oliveira}
\affiliation{Instituto de Física, Universidade Federal do Rio de Janeiro, Av. Athos da Silveira Ramos, 149, Cidade Universitária, Rio de Janeiro, RJ, Brazil}

\author{Santiago Perez}
\email{santiep.137@gmail.com}
\affiliation{Universidad de Buenos Aires, Facultad de Ciencias Exactas y Naturales, Departamento de Física. Buenos Aires, Argentina.}
\affiliation{CONICET - Universidad de Buenos Aires, Instituto de Física de Buenos Aires (IFIBA). Buenos Aires, Argentina}

\author{Laura        Paulucci}
\affiliation{Universidade Federal do ABC, Avenida dos Estados 5001, Santo André, SP, Brazil}

\author{Dario        Rodrigues}
\email{rodriguesfm@df.uba.ar}
\affiliation{Universidad de Buenos Aires, Facultad de Ciencias Exactas y Naturales, Departamento de Física. Buenos Aires, Argentina.}
\affiliation{CONICET - Universidad de Buenos Aires, Instituto de Física de Buenos Aires (IFIBA). Buenos Aires, Argentina}

\author{Youssef       Sarkis}
\affiliation{Instituto de Ciencias Nucleares, Universidad Nacional Autónoma de México,  Circuito Exterior s/n, C.U., CDMX, Mexico}

\author{Ivan Sidelnik}
\affiliation{Departamento de Física de Neutrones, Centro Atómico Bariloche, (CNEA, CONICET), Bariloche, Argentina} 

\author{Miguel     Sofo Haro}
\affiliation{Universidad Nacional de Córdoba, CONICET (IFEG) and CNEA (RA0), Córdoba, Argentina}
          
\author{Diego        Stalder}
\affiliation{Facultad de Ingeniería, Universidad Nacional de Asunción, Campus de la UNA, San Lorenzo, Paraguay}

\author{Javier        Tiffenberg}
\affiliation{Fermi National Accelerator Laboratory, Batavia, IL, United States}

\author{Pedro         Ventura}
\affiliation{Instituto de Física, Universidade Federal do Rio de Janeiro, Av. Athos da Silveira Ramos, 149, Cidade Universitária, Rio de Janeiro, RJ, Brazil}

\collaboration{CONNIE and Atucha-II Collaborations}%\noaffiliation

\date{\today}% 

\begin{abstract}
Millicharged particles, proposed by various extensions of the standard model, can be created in pairs by high-energy photons within nuclear reactors and can interact electromagnetically with electrons in matter. Recently, the existence of a plasmon peak in the interaction cross-section with silicon in the eV range was highlighted as a promising approach to enhance low-energy sensitivities. The CONNIE and Atucha-II reactor neutrino experiments utilize Skipper-CCD sensors, which enable the detection of interactions in the eV range. We present world-leading limits on the charge of millicharged particles within a mass range spanning six orders of magnitude, derived through a comprehensive analysis and the combination of data from both experiments.
\end{abstract}

%\keywords{Suggested keywords}%Use showkeys class option if keyword

\maketitle

%\tableofcontents

Short baseline reactor antineutrino experiments arise as an opportunity to look for beyond the standard model (BSM) particles, such as QCD axions~\cite{Dent, AristizabalSierra:2020}, and other dark matter candidates~\cite{PhysRevD.105.L111101, PROSPECT2021, PROSPECT2021b}.
Among the extensions to the standard model, millicharged particles (mCP) have gained significant attention and are considered highly compelling BSM candidates~\cite{Dobroliubov:1989mr, Magill:2018tbb, Arguelles2021}.
The TEXONO collaboration employed a point-contact 500\,g germanium detector with a low-energy threshold of 300\,eV, positioned 28\,m away from a 2.9\,GW$_{\rm th}$ nuclear reactor, using 124.2/70.3 kg-days of reactor ON/OFF data, to set the most stringent direct laboratory exclusion limits below 1\,MeV up to date~\cite{TEXONO:2018nir}.

Silicon detectors have pushed the energy thresholds to lower values, allowing for higher sensitivity in the dark-matter low-mass range. 
CONNIE~\cite{CONNIE2019,CONNIE2022} was the first experiment to use silicon charge-coupled devices (CCD) at nuclear reactor to look for coherent elastic neutrino-nucleus scattering and impose competitive constraints on BSM physics~\cite{CONNIE2020}.  
Recently, CONNIE upgraded its detector by substituting the CCDs with a pair of Skipper-CCDs~\cite{Tiffenberg:2017aac}, increasing its sensitivity and low-energy reach~\cite{CONNIE2024}. 
At the same time, the Atucha-II experiment has deployed a Skipper-CCD sensor inside a nuclear power plant, situated just 12\,m from the reactor core. The initial results demonstrate promising potential to explore both standard-model physics and exotic searches~\cite{atucha2024}.

In a recent contribution, the SENSEI experiment has achieved the most stringent exclusion limit for mCPs in the 30 to 380\,MeV mass range using Skipper-CCDs~\cite{SENSEI_mCP}. Based on this, additional promising scenarios for enhancing this search have been proposed using a kg-scale experiment with the same technology~\cite{Oscura:2022vmi}. 
There is, however, a compelling reason to conduct such an experiment in the vicinity of a nuclear reactor, as its core represents the most potent source of gamma rays on Earth. 
Moreover, mCPs produced by Compton-like interactions from gamma rays enable the exploration of sub-MeV masses, making the reactor experiments the most competitive in this energy range.

Tracking mCPs passing through a stack of detectors has also been proposed as a highly sensitive strategy for observing these particles when produced in an accelerator~\cite{Oscura:2023qch}. However, this strategy is not competitive below 1\,MeV where 
the fraction of the elementary charge, $\varepsilon$, have already been excluded well beneath to 10$^{-5}$ ~\cite{TEXONO:2018nir}. Observing tracks for even smaller values of $\varepsilon$ would require extremely large stacked detectors.

In this Letter, we present world-leading direct laboratory exclusion limits for mCPs below 1\,MeV, covering six orders of magnitude in mCP mass. 
This work also represents a collaborative effort between the CONNIE and Atucha-II experiments—the first to use Skipper-CCD technology for this type of search in reactor experiments—achieving combined results that enhance the robustness of the analysis and yield a stronger limit.

%%%%%%%%%%%%%%%%%%%%%%%%%%%%%%%%%%%%%%%%%%%%%%%%%%%%%%%%%%%%%%%%%%%%%%%%%%%%%

A possible channel for the generation of mCPs ($\chi_q$) in the sub-GeV mass range is through a Compton-like process, in which a $\gamma$ photon scatters off an electron in the material of the reactor core (Uranium), the photon could then kinetically mix with a dark photon  \cite{Adshead:2022ovo,Feng:2023ubl}. 
As a result, a new dark fermion pair $\chi_q - \overline{\chi}_q$, coupled to the dark photon, can acquire a small electric charge $q=\varepsilon e$ proportional to the kinetic mixing parameter.
In the case of a pure mCP, the dark photon is massless and the resulting particles has an electromagnetic charge $\varepsilon$ which is a real number with $|\varepsilon|<1$ and $e$ is the elementary charge. 

Approximately half of the $\gamma$ flux in the reactor core arises from highly excited fission fragments, with the remainder originating from radioactive de-excitation of the daughter nuclei, inelastic neutron scattering, and capture of neutrons by core materials. 
The $\gamma$-ray spectrum characteristic of neutron-induced fission of uranium, determined for an FRJ-1 (Merlin) research reactor core, can be parameterized by
\begin{equation}
    \frac{dN_{\gamma}}{dE_{\gamma}}=K \, P\, \exp({-1.1E_{\gamma}})  \,,
\label{eq:gamma_flux}
\end{equation}
which holds for photon energies $E_\gamma$ above 0.2\,MeV, where $K=0.581\times 10^{18}$\,MeV$^{-1}$s$^{-1}$, and $P$ is the reactor thermal power in MW. 
This model has been applied by the TEXONO collaboration~\cite{TEXONO:2018nir} and has also been employed in recent dark matter searches~\cite{AristizabalSierra:2020,TEXONO:2018nir,Park:2017prx,Arias-Aragon:2023ehh, Smirnov:2021wgi}.
To the best of our knowledge, however, secondary $\gamma$-rays were not taken into account in previous calculations although they are also able to produce mCPs. In this work, we calculate the limits in both scenarios, reflecting the contributions to mCP production of only primary $\gamma$-rays, as well as including secondary $\gamma$-rays from transport and energy loss in the nuclear core. The secondary contribution was estimated from a simulation using GEANT4~\cite{Agostinelli:2002hh, Allison:2006ve}.

The differential production cross-section for mCP through the mentioned channel can be estimated by adapting the lepton-pair production process~\cite{TEXONO:2018nir}. 
To do this, the $\chi_q-\overline{\chi}_q$ production vertex is parameterized by $\varepsilon$ and the lepton mass is replaced by the proposed mCP mass, $m_{\chi_q}$,

%\begin{widetext}
\begin{eqnarray}
\frac{d\sigma}{dE_{\chi_q}}(\gamma e \rightarrow \chi_q \overline{\chi}_q e) = \frac{4}{3}\frac{\varepsilon^2\alpha^3}{m_e^2E_{\gamma}^3} \times \;\;\;\;\;\;\;\;\;\;\;\;\;\;\;\;\;\;\;\;\;\;\;\;\;\;\;\;\;\; \\ 
 \times \big[(3(E^2_{\chi_q}+E^2_{\overline{\chi}_q })+2E_{\chi_q}E^2_{\overline{\chi}_q}\big] \log\Big(\frac{2E_{\chi_q}E_{\overline{\chi}_q}}{E_{\gamma}m_{\chi_q}}\Big)\,, \nonumber
\label{eq:production}
\end{eqnarray}
%\end{widetext}
%
where $m_e$ is the mass of the electron, $\alpha$ is the fine structure constant, and $E_{\overline{\chi}_q}=E_{\gamma}-E_{\chi_q}$.

The total differential flux of mCP is calculated by taking the convolution of the reactor $\gamma$-ray spectrum and the differential production cross-section as in Eq.~\ref{eq:diff_flux}.
The integral is normalized by the total interaction cross-section, $\sigma_{\rm tot}$. 
Following the same argument as in Refs.~\cite{TEXONO:2018nir, PhysRevD.75.075014}, restricting the integral between 1 and 5 MeV, $\sigma_{\rm tot}$ can be approximated by the Compton cross-section, since this mechanism dominates over other interaction processes. The flux then becomes,
\begin{equation}
\frac{d\phi_{\chi_q}}{dE_{\chi_q}}=\frac{2}{4\pi D^2}\int\frac{1}{\sigma_{\rm tot}}\frac{d\sigma}{dE_{\chi_q}}\frac{dN_{\gamma}}{dE_{\gamma}}dE_{\gamma} \,,
\label{eq:diff_flux}
\end{equation}
where $D$ is the distance between the detector and the center of the reactor core, and there is a factor of 2 from the fact that mCPs are produced in pairs.

\begin{figure}[t]
\centering
\includegraphics[width=.50\textwidth]{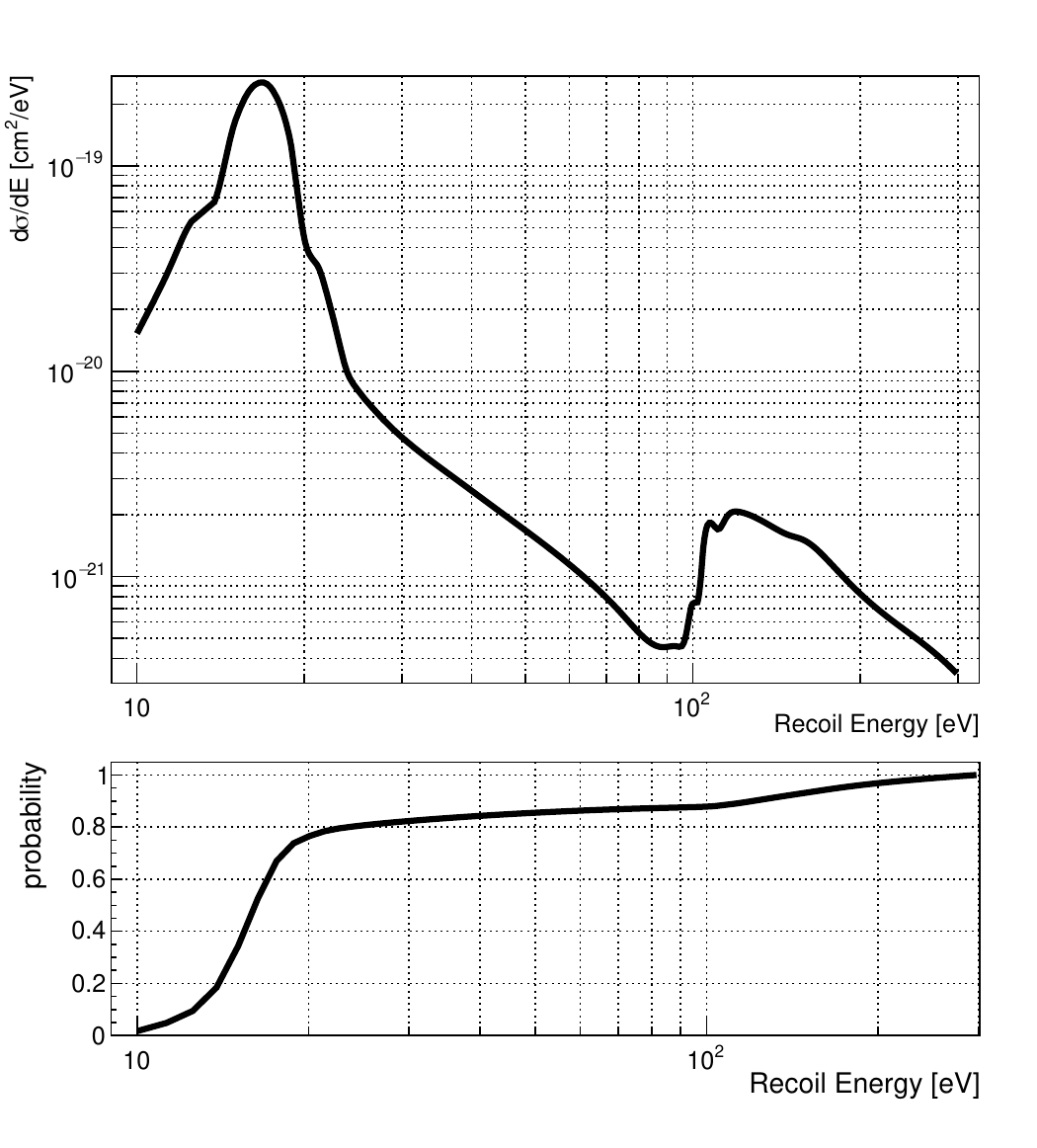}
\caption{(Top) Differential interaction cross-section in silicon for mCP with mass $m_{\chi_q}=1$\,eV and charge fraction $\varepsilon=10^{-6}$, and (bottom) cumulative probability of interaction normalized to the energy interval under consideration.}
\label{fig:cross_section}
\end{figure}

%%%%%%%%%%%%%%%%%%%%%%%%%%%%%%%%%%%%%%%%%%%%%%%%%%%%%%%%%%%

Millicharged particles can be described as charged particles traveling with a velocity $\beta = \frac{p}{E_{\chi_q}}$, and are expected to interact electromagnetically, leading to the production of ionization wherein free charge carriers are released for energy depositions above the band gap of silicon. 
A first description of this phenomenon can be found in Ref.~\cite{PhysRev.57.485}. This description is semiclassical in the sense that the particle is considered to be relativistic with a $p_{\mu}$ and a classical source of electromagnetic fields. The effective cross-section for the interaction is then given by
%
%\begin{widetext}
\begin{eqnarray} 
\label{energy-loss-fermi}
\frac{d\sigma}{d\omega} = \frac{8\alpha\varepsilon^2}{N_e\beta^2}\int_0^\infty dk \bigg\{\frac{1}{k}\mathrm{Im}\left(-\frac{1}{\epsilon(\omega, k)}\right) +  \\
~+k\left(\beta^2 - \frac{\omega^2}{k^2}\right)\mathrm{Im}\left(\frac{1}{-k^2 + \epsilon(\omega, k)\omega^2}\right)\bigg\} \,, \nonumber
\end{eqnarray}
%\end{widetext}
%
where information regarding the interaction between virtual photons emitted by the mCP and the material is encoded in the complex dielectric function $\epsilon(\omega,k)=\epsilon_1(\omega,k)+i\epsilon_2(\omega,k)$, $N_e$ is the electron number density of the material and $k$ is the momentum transfer to the material.

The Photo Absorption Ionization (PAI) model (also known as Fermi virtual photon or Weizsacker-Williams approximation) has been used to describe the energy loss per unit length for standard model particles, but it has also been employed in mCP searches by scaling the coupling with the electron charge to the mCP charge~\cite{SuperCDMS:2020hcc,kamdin2018search}. A full derivation of the modeling of the cross-section can be found in Ref.~\cite{Allison:1980vw}, as well as a discussion on the physical meaning of each term. This differential cross-section can be expressed as
\begin{eqnarray}
\frac{d\sigma}{dE}=&\frac{\alpha}{\beta^2\pi}\frac{\sigma_{\gamma}(E)}{EZ}\ln\big[(1-\beta^2\epsilon_1)^2+\beta^4\epsilon_{2}^2\big]^{-1/2} \nonumber \\
& +\frac{\alpha}{\beta^2\pi}\frac{1}{N_e\hbar c}\Big(\beta^2-\frac{\epsilon_1}{|\epsilon|^2}\Big)\Theta +\frac{\alpha}{\beta^2\pi}\frac{\sigma_{\gamma}(E)}{EZ}\ln\Big(\frac{2mc^2\beta^2}{E}\Big) \nonumber \\
& +\frac{\alpha}{\beta^2\pi}\frac{1}{E^2}\int^E_0\frac{\sigma_{\gamma}(E')}{Z}dE' \,, 
\label{eqn:sigma}
\end{eqnarray} 
where $\sigma_{\gamma}$ is the photoabsorption cross-section 
and $\tan\Theta=\epsilon_2\beta^2/(1-\beta^2\epsilon_1)$. 

Figure~\ref{fig:cross_section} illustrates the energy dependence of the interaction cross-section for an mCP with 1\,eV mass and $\varepsilon=10^{-6}$. 
The sizable enhancement observed in the low-energy region comes from the interaction of the particles with bulk plasmons. 
The tabulated complex index of refraction and photoabsorption data is not valid for energy deposits below $\sim50$\,eV as there is a crucial difference between optical absorption and the scattering of relativistic particles. Photons are always transversely polarized, whereas a charged particle can also interact with the material via longitudinal Coulomb modes which dominate the response function in this regime. Optical absorption data have then been identified as a poor proxy for a relativistic scattering of charged particles at low energies. 

To precisely calculate the interaction cross-section for mCPs near the plasmon peak, in Ref.~\cite{Essig:2024ebk} the electron loss function Im$(-1/\epsilon(\omega,k))$ is calculated for different models of the dielectric function in the DarkELF package~\cite{DarkELF}, producing reliable results when compared to electron energy loss spectroscopy data when $\beta\geq 0.01$. 

Thus, we turn to the DarkELF (GPAW) model to calculate the expected rate of mCPs below 50\,eV as PAI model underestimates the cross-section near the plasmon peak, while GPAW provides a more accurate description, as stated in Ref.~\cite{Essig:2024ebk}. This integration constitutes the main contribution to systematic uncertainty (see Supplemental Material~\cite{supp} for details).

The differential rate of events due to $\chi_q$ interactions can then be calculated by integrating the flux obtained in Eq.~\ref{eq:diff_flux} convolved with the interaction cross-section,
\begin{equation}
    \frac{dR}{dE}=\rho\int_{E_{\rm min}}^{E_{\rm max}}\frac{d\phi_{\chi_q}}{dE_{\chi_q}}\frac{d\sigma}{dE}dE_{\chi_q} \,,
    \label{eq:cross_section}
\end{equation}
where $\rho$ is the atomic number density and ($E_{\rm min}$, $E_{\rm max}$) are the values of the mCP energy.

%%%%%%%%%%%%%%%%%%%%%%%%%%%%%%%%%%%%%%%%%%%%%%%%%%%%%%

Fully depleted high-resistivity silicon CCDs are being used in various experiments dedicated to dark matter~\cite{DAMIC:2020cut} and low-energy neutrino detection, which require low thresholds and excellent background control. Skipper-CCDs~\cite{Tiffenberg:2017aac} are the new generation of this imaging technology, achieving single-electron sensitivity by using the non-destructive readout of the charge packets held in each pixel~\cite{Arnquist_2023, DAMIC:2023ela, Oscura:2023qch, SENSEI:2023zdf}. 
This capability allows for an excellent signal-to-noise ratio with detection thresholds as low as 15\,eV in above-ground experiments~\cite{Moroni2022}.
Both the CONNIE and Atucha-II experiments employ Skipper-CCD sensors of 675\,$\mu$m thickness and 15\,$\mu$m pixel size, designed by LBNL Microsystems Laboratory and manufactured by Teledyne-DALSA. They are read out using a Low-Threshold Acquisition (LTA) controller board~\cite{Cancelo2021}.

The Coherent Neutrino-Nucleus Interaction Experiment (CONNIE)~\cite{CONNIE2019, CONNIE2022, CONNIE2020, CONNIE2024} is operating in a ground-level laboratory outside the dome of the 3.95\,GW$_{\rm th}$ Angra 2 nuclear reactor near Rio de Janeiro, Brazil, at a distance of about 30\,m from the core. The experiment has been taking data with two Skipper-CCDs of 0.247\,g mass each since July 2021, totaling an exposure of 18.4\,g-days. 
The sensors achieved an ultra-low readout noise of 0.15\,e$^-$ by sampling 400 times each pixel charge, allowing to measure for the first time the energy spectrum near a reactor down to a threshold of 15\,eV\@, and obtaining a background rate in reactor-OFF data of around 4\,$\rm{kg^{-1}d^{-1}keV^{-1}}$~\cite{CONNIE2024}. 

The Atucha-II nuclear power plant is a 2.175\,GW$_{\rm th}$ pressurized heavy water reactor that utilizes natural UO$_{2}$ as fuel, located in Buenos Aires province, Argentina. Since December 2021, a Skipper-CCD with 2.2\,g mass is taking data inside the containment sphere, 12\,m away from the nuclear core. The sensor is operated with a readout noise of 0.17\,e$^{-}$, achieved by averaging 300 samples of the charge in each pixel, and a background rate in reactor-OFF data of around 30\,$\rm{kg^{-1}d^{-1}keV^{-1}}$.
The dataset used in this work corresponds to 82\,g-days of unpublished data from the 2023 run. 

In the interaction cross-section from Fig.~\ref{fig:cross_section}, the plasmon peak located between 10 and 25\,eV is easily identified as the most convenient region to look for mCPs.
Above 25\,eV, the cross-section decreases significantly until reaching 100\,eV\@. At this point, the energy released in the detector is sufficient to ionize the silicon $p$ shell, increasing in 6 the number of electrons acting as targets. 
Based on this and before unblinding the data, a 200\,eV energy interval starting at the lowest energy possible in each experiment was established (see Table~\ref{tab:results}). 
It is noteworthy that the CONNIE threshold of 15\,eV enables the inclusion of most of the plasmon peak while extending the interval above 240\,eV for Atucha II would result in a loss of sensitivity due to the decrease in the cross-section against a constant background rate.

\begin{table}[b!]
\centering
\caption{Experimental observables of each experiment. }
\label{tab:results}
\begin{ruledtabular}
\begin{tabular}{lcc}
%\hline \hline
Observable    & CONNIE & Atucha-II \\ \hline
Reactor ON exposure {[}g-day{]}   & 14.9    & 59.4     \\ 
Reactor OFF exposure {[}g-day{]}    & 3.5  & 22.6    \\ 
Energy bin {[}eV{]}     & 15--215   & 40--240   \\ 
Reactor ON counts     & 6    & 168    \\ 
Reactor OFF counts     & 2      & 71   \\ 
90\% C.L. upper limit on events   & 6.2  & 30.9   \\ %\hline
\end{tabular}
\end{ruledtabular}
\end{table}

Table~\ref{tab:results} summarizes the results obtained from each experimental run. Upper limits at 90\% C.L. on the numbers of events were computed using the frequentist approach, as outlined in Ref.~\cite{Cowan2011}. 
Efficiency corrections were implemented by convolving the expected theoretical event count with the efficiency curves from Refs.~\cite{CONNIE2024, atucha2024}.

\begin{figure*}[t]
\centering
\begin{minipage}{0.49\textwidth}
    \includegraphics[width=\textwidth]{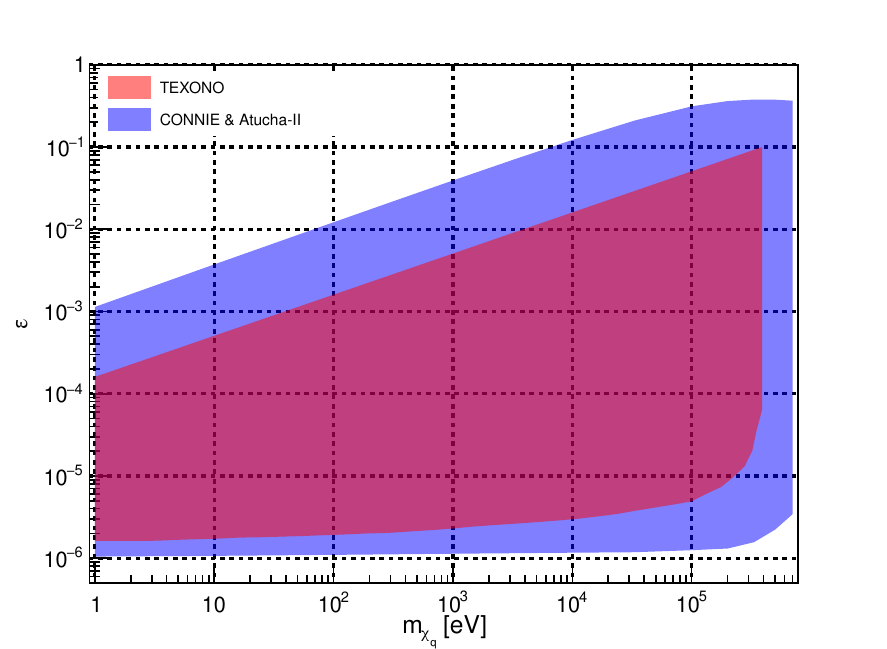}
\end{minipage}\hfill
\begin{minipage}{0.49\textwidth}
    \includegraphics[width=\textwidth]{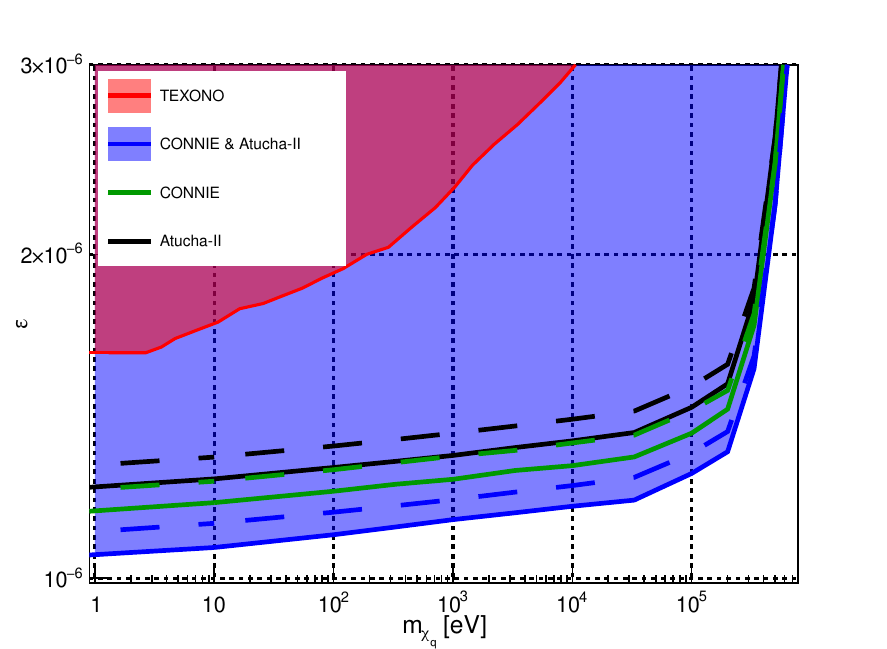}
\end{minipage}
\caption{Exclusion limits at 90\% C.L. as a function of the mCP mass and charge fraction for the CONNIE (green), Atucha-II (black), and their combination (blue). The TEXONO exclusion limit~\cite{TEXONO:2018nir} is also included for comparison (red). 
\textbf{Left:} Upper limits are derived from the attenuation of mCPs as they travel from the reactor to the detector according to Ref.\cite{Arefyeva2022}. 
\textbf{Right:} Solid lines represent results considering mCP production from both primary and secondary $\gamma$-rays, while dashed lines account for only primary $\gamma$-rays.}

\label{fig:limits}
\end{figure*}

Figure~\ref{fig:limits} depicts the independent exclusion limits attained by each experiment. 
The results show an improvement by the CONNIE and Atucha-II experiments for two cases: primary plus secondary (solid lines), or only primary $\gamma$-ray production (dashed lines). 
Moreover, a combined analysis of the outcomes from both experiments is performed following Ref.~\cite{Cowan2011}, yielding a further improvement on the individual results (see Supplemental Material~\cite{supp} for details).
The upper limit was derived based on the attenuation of mCPs in the material between the reactor and the detector, accounting for Bremsstrahlung and ionization stopping power, as outlined in Ref.~\cite{Arefyeva2022}
The total combined error obtained through Monte Carlo propagation was 9.5\%. 
A detailed description of each source and its contribution can also be found in the Supplemental Material~\cite{supp}.

%%%%%%%%%%%%%%%%%%%%%%%%%%%%%%%%%%%%%%%%%%%%%%%%%%%%%%%%%%%%%%%

The mCP search is particularly challenging because the number of observed events scales with $\varepsilon^4$. A factor $\varepsilon^2$ originates from the mCP production within the reactor core (Eq.~\ref{eq:production}), while another factor $\varepsilon^2$ results from the probability of interaction (Eq.~\ref{energy-loss-fermi}). 
Therefore, achieving a 2.5-fold improvement in the limit around 100 keV (see Fig.~\ref{fig:limits}) is equivalent to a 40-fold increase in experimental exposure. Such an enhancement is impractical with previous strategies, given that the best results to date was achieved with a half-kilogram detector and exposure times of hundreds of days~\cite{TEXONO:2018nir}.
On the other hand, mCP flux scales linearly with the thermal power of the reactor but inversely with the square of the distance. Achieving such an increase in flux would require reducing the distance by a factor of six, which is unfeasible in most reactor experimental scenarios.

%%%%%%%%%%%%%%%%%%%%%%%%%%%%%%%%%%%%%%%%%%%%%%%%%%%%%%%

The capability of observing interactions at the eV scale allows CONNIE to take advantage of the plasmon resonance~\cite{liang2024plasmonenhanced}.  
This collective mode of electronic excitation in semiconductor materials significantly enhances the detection sensitivity for sub-MeV relativistic particles.
Accessing recoil energies in the eV range results in a flatter cross-section dependence on mCP mass, compared to the TEXONO experiment, which operated above 300 eV, showing a logarithmic dependence on the inverse of the mCP mass through the equivalent photon approximation cross-section~\cite{TEXONO:2018nir}.
The Atucha-II experiment possesses a three times higher mCP flux and a four times higher reactor ON exposure compared to the CONNIE experiment.
This is compensated by the fact that the CONNIE's energy interval accounts for a 4.5 times larger probability of mCP interaction (see Fig.~\ref{fig:cross_section}) than Atucha's energy interval, as well as a background rate six times lower.
Consequently, both experiments feature a similar signal-to-square-root-of-background ratio, which explains the closeness of the achieved exclusion limits.

We have demonstrated an alternative approach to improving limits using Skipper-CCD technology, which combines a 1.1 eV silicon band gap with sub-electron readout noise. This enables experiments with sensitivity at the eV scale to exploit the significant enhancement in the cross-section.
We present for the first time a combined analysis of the collaborative effort between the two experiments, resulting in a very robust limit that strongly mitigates systematic errors eventually introduced by any of them. 
The combined result extends the experimental exclusion limits further towards 700\,keV in mCP mass, as well as down towards $\varepsilon\sim1\times 10^{-6}$ for the lowest mCP masses of 1\,eV, thus becoming the best direct laboratory constraint on mCP coupling.
Futhermore, both experiments are currently conducting this search with only around 1\,g of sensor, and have plans for substantial increases in sensor mass~\cite{atucha2024, CONNIE2024}. The scaling of mass to reach the kilogram range has already been developed~\cite{Oscura:2022vmi, Oscura:2023qch}, and the potential physics impact of such an experiment has also been evaluated~\cite{VIOLETA2021, VIOLETA2022}.

\begin{acknowledgments}
CONNIE thanks the Silicon Detector Facility staff at the Fermi National Accelerator Laboratory for hosting the assembly and test of the detector components used in the CONNIE experiment. The CCD development was partly supported by the Office of Science, of the U.S. Department of Energy under Contract No.~DE-AC02-05CH11231. 
We are grateful to Eletrobras Eletronuclear, and especially to G. Coelho, I. Soares, I. Ottoni and L. Werneck, for access to the Angra 2 reactor site, infrastructure and the support of their personnel. We express gratitude to R. Shellard (in memoriam) for supporting the experiment. We thank M. Giovani for the IT support and M. Martínez Montero for the technical assistance. 
We acknowledge the support from the Brazilian Ministry for Science, Technology, and Innovation and the funding agencies FAPERJ, 
%(grants E-26/110.145/2013, E-26/210.151/2016, E-26/010.002216/2019, E-26/202.687/2019, E-26/210.079/2020), 
CNPq 
%(grants 437353/2018-4, 407707/2021-2), 
and FINEP; 
%(RENAFAE grant 01.10.0462.00); 
Argentina's CONICET and AGENCIA I+D+i; (grants PICT-2019-2019-04173; PICT-2021-GRF-TII-00458; PICT-2021-GRF-TI-00816) and the LAA-HECAP Network.
We acknowledge the support from Mexico’s CONAHCYT (grant CF-2023-I-1169) and DGAPA-UNAM (PAPIIT grant IN104723).
This work made use of the CHE cluster, managed and funded by COSMO/CBPF/MCTI, with financial support from FINEP and FAPERJ, and operating at the Javier Magnin Computing Center/CBPF.

Atucha-II thanks the NA-SA team in Argentina for all the support during the deployment and operation of the Skipper-CCD system at Atucha-II. This work was supported by Fermilab under DOE Contract No. DE-AC02-07CH11359. We thank  Eneas Kapou from NA-SA for assisting in data transportation outside the plant. We thank Daniel Cartelli from CNEA for creating the artistic plots of the plant and the system. We also want to thank Eduardo Arostegui who insisted on carrying on with the project and facilitated a lot the communication with the plant during the first phase of the deployment.
\end{acknowledgments}

\bibliography{main} 

%\clearpage
\section{Supplemental Material}

\subsection{Nuclear reactor $\gamma$-ray spectrum}

The photon flux and photon spectrum typically used in the calculation for the production of new particle candidates are approximations of the fission prompt $\gamma$-rays or delayed $\gamma$-rays from fission products. For example, the spectrum produced in Eq.~\ref{eq:gamma_flux} is the photon production resulting from neutron capture in $^{235}$U\@. Most analyses do not consider the transport and energy loss of these photons in the reactor core as an additional production probability for generating the particles of interest. Since the photons lose energy due to the successive scattering, this additional probability of production of exotic particles will favor those with lower energy. 

\begin{figure}[h]
\centering
\includegraphics[width=.49\textwidth]{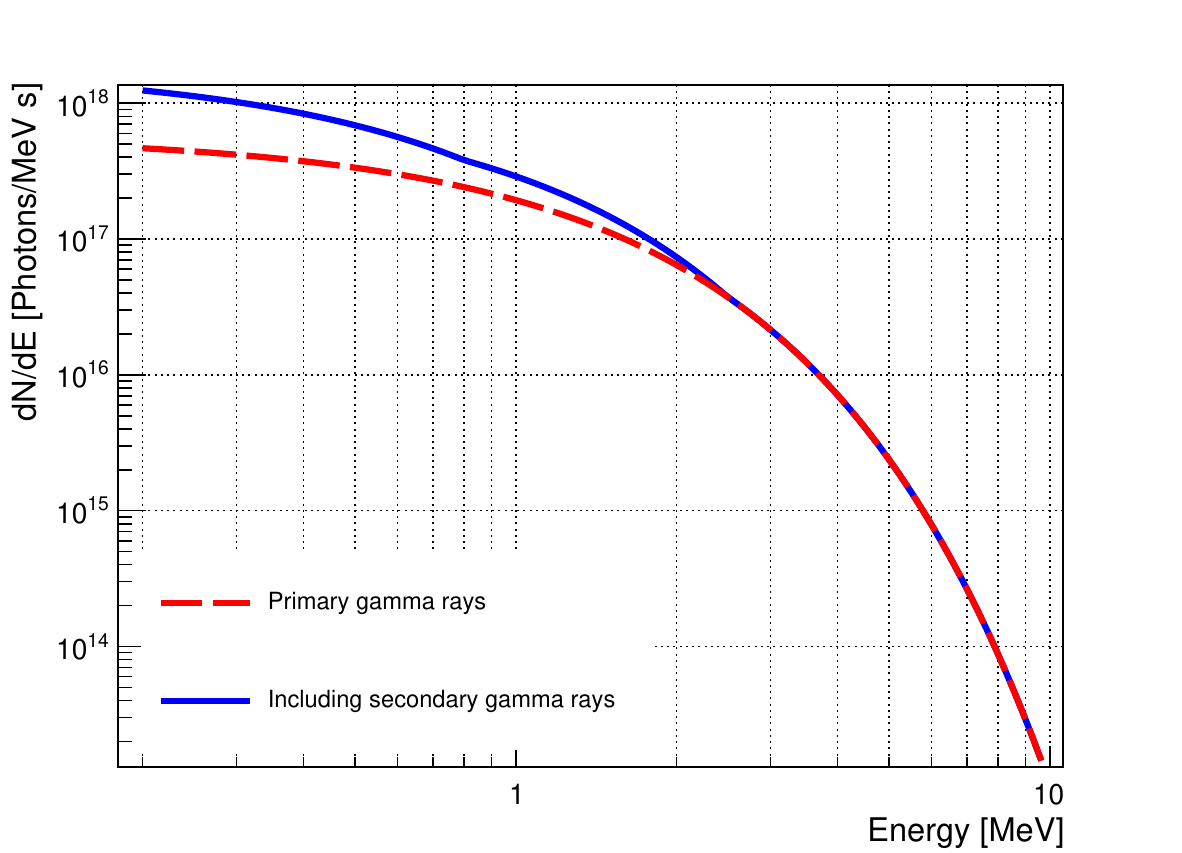}
\caption{Photon flux produced in the nuclear reactor. The red dashed line is derived from Eq.~\ref{eq:gamma_flux}, considering primary $\gamma$-rays only. The blue solid line is obtained from a GEANT4 simulation, taking into account secondary photon production.}
\label{fig:gamma_flux}
\end{figure}

A simulation was performed using GEANT4~\cite{Agostinelli:2002hh, Allison:2006ve} to transport primary photons produced in the core. A single-volume geometry of the uranium element was used to track the successive interactions of $\gamma$ particles. The particle source of $\gamma$-rays follows Eq.~\ref{eq:gamma_flux} generated isotropically in the center of the volume. The interactions of the $\gamma$-rays are tracked through the volume, with the energy before each interaction recorded as a candidate for interactions in the material core. Figure~\ref{fig:gamma_flux} shows the original primary $\gamma$ spectrum and the one including secondary photons from the simulation of interacting prompt photons. The additional secondary photons are more prominent in the low-energy part of the spectrum. 

\subsection{Statistical formalism}

To obtain the upper limit of $\varepsilon$, we followed the approach of Ref.~\cite{Cowan2011}, based on computing the likelihood profile ratio. The two measured random variables in our experiments are the detected number of events in a given energy interval for the reactor-ON ($n$) and reactor-OFF ($m$) runs. Thus, the likelihood function of each experiment can be expressed as the product of the reactor-ON and the reactor-OFF Poisson distributions: 
\begin{center}
$L(\mu, b) = \frac{(\mu s + b)^n}{n!}e^{-(\mu s + b)} \times \frac{(\tau b)^m}{m!}e^{-(\tau b)}$ \,,
\end{center}
where $\mu$ denotes the strength parameter, $s$ is the expected number of counts for a given charge fraction $\varepsilon_0$ in that energy interval (given by the theoretical prediction),
$b$ is the mean number of expected background events in the reactor-ON data, and $\tau$ is a factor that corrects $b$ to the exposure of reactor OFF: $\tau = \frac{\text{exposure OFF}}{\text{exposure ON}}$. 

To determine the upper limit of $\mu$, we first compute the profile likelihood ratio and then the statistic $t_{\mu}$: 
\begin{center}
$\lambda \equiv \frac{L(\mu, \hat{\hat{b}})}{L(\hat{\mu},\hat{b})} \qquad \text{and} \qquad t_{\mu} \equiv -2\ln (\lambda)$\,.
\end{center} 
Here $\hat{\mu}$ and $\hat{b}$ are the maximum likelihood estimators and $\hat{\hat{b}}$ is the result of maximizing the parameter $b$ for a given value of $\mu$. Usually, $b$ is called a \textit{nuisance parameter}. The statistic $t_{\mu}$ has a chi-squared probability distribution with 1 degree of freedom ($f(t_{\mu} \vert \mu)$). This can be used to compute the p-value of the particular experimental realization for a given hypothesis $\mu$: 
\begin{center}
p-value $ =\int _{t_{\mu , obs}} ^{\infty} f(t_{\mu} \vert \mu) dt_{\mu}$\,.
\end{center}

The upper limit of $\mu$ with a $\alpha$\% confidence level corresponds to the value of this parameter that gives a p-value $=1 - \alpha$. Then, the upper limit of $\varepsilon$ is calculated as $\sqrt[4]{\mu}\,\varepsilon_0$.

This procedure can be applied for each experiment separately but also represents a useful and robust framework for obtaining the upper limit of a combination of the two experiments. In this case, the total experiment has four measured random variables $(n_a, m_a, n_c, m_c)$ where the $a$ and $c$ indices correspond to Atucha-II and CONNIE, respectively. 
Since the two experiments are independent, the new combined likelihood is the product of the two single likelihoods: 
 \begin{center}
$ \displaystyle L(\mu, b_a, b_c) = \prod_{i = a,c}\frac{(\mu s_i + b_i)^{n_i}}{n_i!}e^{-(\mu s_i + b_i)} \times \frac{(\tau b_i)^{m_i}}{m_i!}e^{-(\tau b_i)}$ \,.
\end{center}

Here all $b_i$ are nuisance parameters. The likelihood profile ratio translates into $\lambda = \frac{L(\mu, \hat{\hat{b_a}}, \hat{\hat{b_c}})}{L(\hat{\mu},\hat{b_a}, \hat{b_c})}$. The expected number of events $s_i$ has to be calculated for each experiment because the energy intervals could differ, but the rest of the procedure remains the same. 
The combination of the two experiments shows a more restricted $\alpha$ confidence level interval since it incorporates both realizations as evidence of the absence of millicharged particles.

\subsection{Systematic errors}

Table~\ref{tab:uncertatiny} summarizes the impact of each source of systematic error on the uncertainty of the exclusion limit. To determine the contribution of each source on $\varepsilon$, uncertainty propagation was conducted via toy Monte Carlo. In all cases, the uncertainty in the source is mitigated due to the $\varepsilon^4$ dependence.

The primary contribution stems from matching the DarkELF (GPAW) and PAI cross-sections at 50 eV. We take the average of both models to establish the limit. Modeling its error as a uniform random variable with a relative standard deviation of 19\%, we safely cover the full range between the DarkELF (GPAW) and PAI cross-section values at 50 eV. Although the latter represents the most discrepant point, we consider this uncertainty conservatively across the entire range. As a result, $\varepsilon$ exhibits fluctuations with a standard deviation equal to 4.9\% of its nominal value.

\begin{table}[h!]
    \centering
    \caption{Systematic uncertainties from each source and its propagation on the limit.}
    \label{tab:uncertatiny}
    \begin{tabular}{@{}lcc@{}}
        \toprule
        \textbf{Source} & \textbf{Uncertainty [\%]} & \textbf{Error on limit [\%]} \\
        \midrule
        cross-section & 19             & 4.9             \\
        mCP flux       & 10            & 2.5             \\
        energy resolution              &$<$8     & 7.7     \\
        efficiency curve               &2      & 0.5     \\
        \midrule
        \textbf{Total:} &               & 9.5             \\
        \bottomrule
    \end{tabular}

\end{table}

Contributions from the primary gamma-ray flux, as well as secondary gamma rays calculated using GEANT4 simulation, contribute to the limit through the resulting uncertainty in the mCP flux, which was conservatively estimated at 10\%.

Thanks to the sub-electron readout noise of our measurements, of 0.16~e$^-$ in average, we can accurately resolve the number of electrons in each event~\cite{Rodrigues:2020xpt}. If all the charge of the event falls in only one pixel, the uncertainty equals 0.6~eV. This is the readout noise multiplied by 3.75~eV the energy required on average to ionize one electron~\cite{Rodrigues:2020xpt}.
The worst-case scenario occurs at the lowest energy events of 15~eV (equivalent to four ionized electrons), when the charge is distributed as four electrons across four adjacent pixels. In this case, the cluster readout noise corresponds to 0.32~e$^-$, which translates to an energy uncertainty of 1.2~eV (8\%). This uncertainty decreases significantly as the total charge in the events increases since the charges are summed prior to classification~\cite{Rodrigues2023}. Nevertheless, we have accounted for this impact by integrating the cross-section with a 1.2~eV shift applied to the limits of the energy range. As a result, the interaction probability can be reduced up to a 26\%. Following again a conservative approach, this maximum value was used to derive the contribution on $\varepsilon$ using Montecarlo propagation, resulting in a 7.7\%.

The uncertainty in the efficiency curve is derived from fluctuations in the Monte Carlo simulation performed to derive its energy dependence. It is obtained by applying a diffusion model that describes how electrons generated in the silicon bulk propagate toward the CCD surface~\cite{SofoHaro:2019ewy, atucha2024}.
Finally, a total combined uncertainty of 9.5\% was obtained by adding in quadrature the contributions from each source.

\end{document}